# Impurities, or dopants, that is the question

Baptiste Gault[1,2,*], Leonardo Shoji Aota[1], Mathias Krämer[1], Se-Ho Kim[1,3,*]

[1] Max-Planck-Institut für Eisenforschung, Düsseldorf, Germany.

[2] Department of Materials, Royal School of Mines, Imperial College London, London, UK

[3] Department of Materials Science and Engineering, Korea University, Seoul 02841, Republic of Korea

[*] Corr. Authors: sehonetkr@korea.ac.kr, b.gault@mpie.de

# Abstract

The numerous stories around LK-99 as a possible room-temperature superconductor over the summer of 2023 epitomise that materials are more than a bulk crystallographic structure or an expected composition. Like all materials, those at the core of technologies for the energy generation transition, including batteries, catalysts or quantum materials draw their properties from a hierarchy of microstructural features where impurities can dramatically influence the outcomes. As we move towards a circular economy, the recycling of materials will inevitably create fluxes of increasingly impure materials, generating new challenges for fabricating materials with controlled properties. Here, we provide our perspective on how high-end microscopy and microanalysis have helped us to understand relationships between synthesis, processing and microstructure, avoiding imprecise or even erroneous interpretations on the origins of the properties from a range of materials. We highlight examples of how unexpected impurities and their spatial distribution on the nanoscale can be turned into an advantage to define pathways for synthesis of materials with new and novel sets of physical properties.

# 1 Introduction

Room-temperature superconductivity may never be achieved, but the publication of a preprint in 2023 put it back in the spotlights [1]. A take-home message from the global effort to reproduce these results is that the composition, structure, and microstructure of a material may well matter more to explain its properties than an expected crystal structure. It is also a reminder that materials science is the science of defects: imperfections in a material's structure often underpin its true physical properties. Magnets are another famous example: the coercivity of the strongest permanent magnets, *e.g.* Sm-Co and NdFeB [2,3], currently only reach 20-30% of the theoretical limit [4] – an issue referred to as Brown's paradox [5]. Why? Because of imperfections in the materials' structure – which include defects in the crystalline structure [6,7] such as dislocation and grain boundaries [8], along with foreign chemical species that occupy interstitial sites in the crystal structure, or substitute atoms on the lattice, or the formation of secondary phases [9].

Collectively, structural and chemical defects form what is often referred to as the microstructure. The microstructure develops based on the chemistry of the system, *i.e.* its composition, which leads the mixture of atoms to physically adopt an organisation that satisfies both thermodynamic and kinetic constraints. The chemistry and physics of a material system are not independent from each other, and ultimately combine to provide the material with a given set of properties that make them attractive for one or more applications. Engineering the electrical properties in semiconductor technology is achieved through the controlled introduction of a low amount of chemical species, *i.e.* dopants, that are physically integrated within the lattice to change the charge carrier density. Materials science can hence be seen additively: it is chemistry and physics and engineering – please note that we are unlikely to be first to say this.

Decades of research in metallurgy have shown clearly that mixing elements in certain proportions does not guarantee that the end-product has the expected composition or phase distribution, and the microstructure is influenced also by *e.g.* kinetics and structural defects, and their segregations states [10–12]. These cannot be revealed through bulk characterization of the crystal structure of a solid, and in principle, properties hence cannot be rationalised or readily predicted without knowledge of the microstructure. Processing-microstructure-property are hence necessary for guiding the design of future materials instead of relying on empiricism.

Even the recently reported high-throughput automated synthesis and characterization as part of so-called "autonomous labs" [13], piloted by artificial intelligence [14], target discovery based on data to optimise output, with respect to specific targets, and do not rely on scientific understanding of mechanisms. Their data being imperfect, or imperfectly interpreted at this stage, has led to recent controversies [15], and is unable to capture the true complexity of a material and the relationships with properties – this does not mean that these frontier experimental work will not lead to important advances, but they do need much refinement and complementary work by more highly-resolved techniques.

Overall, these general considerations imply that assessing the microstructure of a material at all stages of its synthesis and processing, as well as over the course of its operation in service, is crucial to understand its properties, their evolution and potentially predict its lifetime. This is however not systematically done, in part because of the challenges associated to performing these measurements, as well as because of cultural differences across communities – once again the report of room-temperature superconductivity in LK-99 is a perfect example. The observed Meissner effect turned out to be likely associated to iron-rich particles and/or the change in conductivity from copper sulfide inclusions that had not been reported due to insufficiently precise characterisation of the investigated material [16]. For active catalysts, this has led to numerous controversies because metallic impurities from synthesis were making organic compounds active, which had been interpreted as coming from the the compounds themselves [17].

In addition, the level of impurities that will be integrated during synthesis of materials is bound to keep increasing in the coming decades, as the fraction of materials from recycling increases in the feedstocks. This is a step towards a circular economy, which is necessary to support the general effort towards reducing the carbon footprint of the material industry – which represent approximately 8% of the global energy use, and nearly 30% of industrial $CO_2$-emissions [18]. This poses new challenges as well for the design and control of materials properties, as minute amounts of a species uncontrollably introduced into a material can dramatically change their properties.

In a recent *viewpoint* article entitled 'Electrocatalysis Goes Nuts', Andrew R. Akbashev [19] argues in the same direction, stating that "*the chemical and structural complexity of electrocatalytic materials and electrodes reported in numerous studies significantly exceeds the quality of their characterization, making meaningful interpretation of the data impossible*". Recent developments in microscopy and microanalysis, enabling composition and structural characterization from the near-atomic up to sub-millimetre scales, must be deployed to actually quantify the composition, structure, their interrelations and their combined influence on the physical properties of interest.

In this perspective article, we draw examples from our recent exploration in a range of nanocatalysts developed for their catalytic properties and discuss how similar concepts apply to other materials systems developed for application in the broad energy context. Our approach typically involves the complementarity between scanning and transmission electron microscopy (S-TEM) and related techniques, such as electron-backscatter diffraction (EBSD) for grain orientation mapping, electron channelling contrast imaging (ECCI) for imaging structural defects on large areas, atomically-resolved STEM and associated X-ray and electron spectroscopies for insights on local atomic organisation and composition, along with atom probe tomography (APT) that provides three-dimensional compositional mapping of materials with sub-nanometre resolution [20]. Here, we showcase how it helps evidencing that the microstructure of materials is simply more complex than expected, and failing to account for this level of complexity hinders accurately interpreting the origins of the properties or their evolution. Embracing the concept of microstructure will open new routes for designing materials with novel combinations of properties.

## 2  Origins of impurities

Fabrication of mono-dispersed colloidal nanomaterials is commonly done by wet-chemical synthesis as it allows to control the size and shape of nanoparticles. The method typically involves three main chemicals: a solvent, a precursor molecule containing the constituent of the nanomaterial, and a reducing agent. By precisely mixing these ingredients and controlling the reaction rate, we obtain colloids with defined morphologies. However, during wet-chemical synthesis, there is always the potential for undesired or uncontrolled incorporation of species invariably present in the solution. This issue is sometimes compounded by the inherent impurities in chemicals. Even high-purity precursor chemicals, as specified in their data sheets, contain notable amounts of unidentified impurities. Recent work by Pan et al. has demonstrated that unwanted Raman peaks, which could lead to misinterpretation, were found in both organic and inorganic experiments due to the interference of ultra-trace impurities in commercial reagents [21].

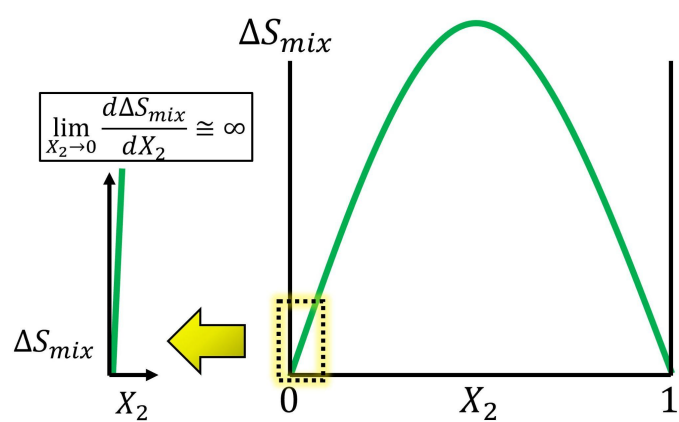

**Figure 1.** Thermodynamics of a minor element (impurity) favourably ingress in a bulk.

Undoubtedly, obtaining chemical reagents with 100% purity is impossible. In addition, from a thermodynamic perspective, the incorporation of secondary element into the system is not surprising considering configuration entropy (see illustration Figure 1). Mixing two elements typically lowers a system's energy as the entropy increases, a concept explained by the entropy of mixing, $\Delta S_{mix}$. According to Gibbs' Theorem, $\Delta S_{mix} = -k (X_1 \ln X_1 + X_2 \ln X_2)$, where k is the Boltzmann constant, and $X_1$ and $X_2$ are the concentrations of the solvent and solute, respectively. Intriguingly, as the concentration of impurities approaches zero, *i.e.* as $X_2$ approaches zero, the mixing entropy's slope becomes steep. This implies that even a trace amount of impurity in a pure system can significantly increase the mixing entropy, thereby stabilizing the system by lowering its Gibbs' free energy of mixing as $\Delta G_{mix} = \Delta H_{mix} - T\Delta S_{mix}$.

Whether these impurities end up in the bulk or on the surface of the nanoparticles will also depend on the free energy gain. For instance, recent research by Kim et al. observed segregation of alkali(-earth) elements on $BaTiO_3$ nanoparticle surface while Sr and Sc elements are uniformly distributed within $BaTiO_3$ nanoparticles [22]. These particles were from commercial product for industrial and laboratory applications that are 99% pure, containing "metal traces". Both the nature and spatial distribution of these traces can have a significant influence on the properties of the nanomaterials [23].

Ultimately, this example highlights the crucial importance of the identification and precise analysis of impurities and their distribution, which remains a highly challenging task [24]. For bulk materials, as exemplified in Figure 2 for metal, semiconductor, insulator, and lab-grown mineral, APT, in conjunction with TEM, has played an increasingly important role for quantifying these impurities in correlation with their atomic-scale distribution across the material's microstructure.

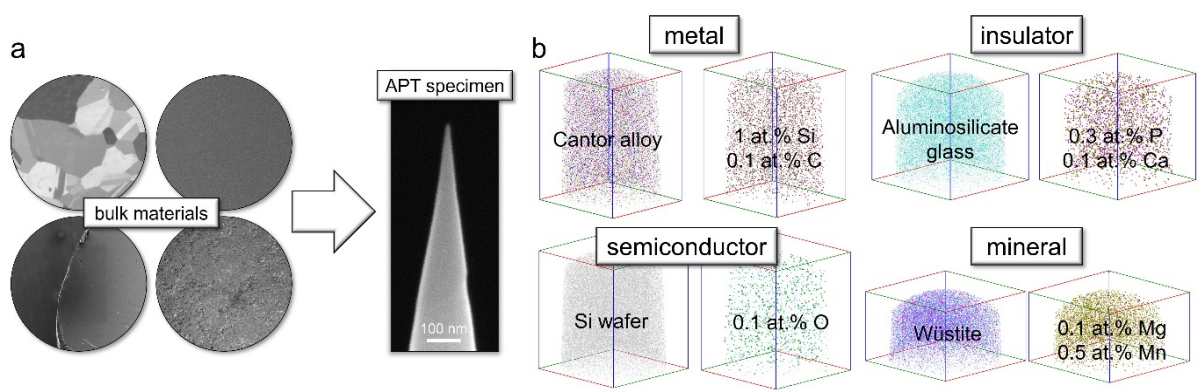

**Figure 2.** Impurities in bulk materials. (a) Scanning electron microscopy (SEM) images of metal, semiconductor, insulator, and mineral and specimens for advanced characterization. (b) 3D atom map of high-entropy Cantor alloy, commercial Si wafer, lab-grown single crystal wustite, and aluminosilicate glass, respectively, containing significant levels of impurities.

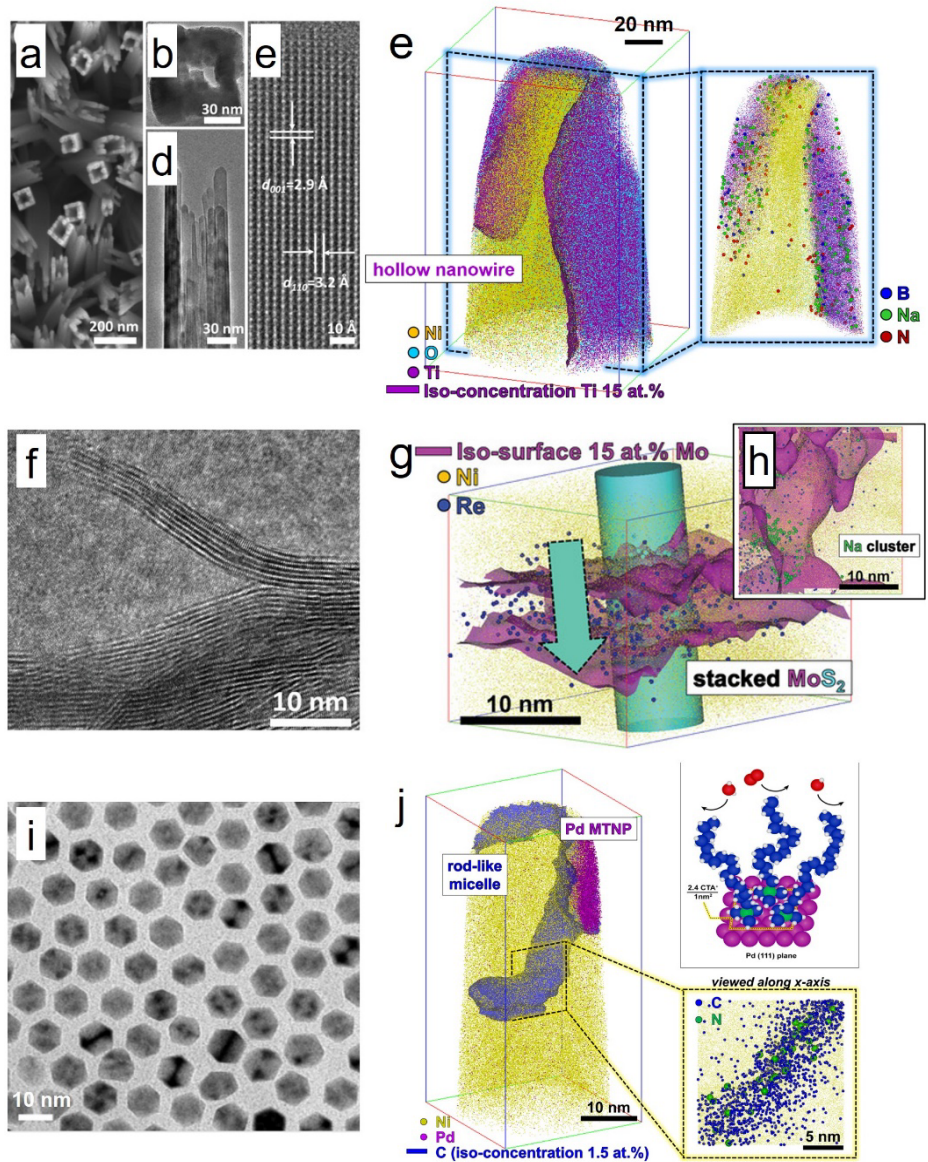

**Figure 2.** Impurities in nano-materials. Electron microscopy images and atom probe reconstructed atom maps of reduced TiO$_2$ hollow nanowire, multiple-twinned Pd nanoparticle, stacked MoS$_2$ nanosheet, respectively. (individual subfigures are from Ref. [25–29]).

# 3 Nanocatalysts

## 3.1 First observations

As the size of the material decreases to the nano-scale, the incorporation of impurities becomes more significant, with their concentration increasing (*e.g.* Gibbs-Thomson effect) [30]. For example, in the case of hydrogen generation applications using reduced $TiO_2$ hollow nanowires synthesized through a wet-chemical method, contradictions in reports of functional properties have arisen due to the lack of precise and quantitative characterization of potential impurities that could play an important role in electro-chemical catalysis as a dopant. Different amounts of trace elements might be introduced during synthesis and have, until now, often been overlooked. Lim et al. [27] have investigated impurity distribution using advanced characterization tool (see Figure 2a-2e) and revealed that unexpected impurities (Na, N, B) were segregated on the surface, which have not been studied well on these to further understand the functional properties of hollow $TiO_2$ nanowires. This was enabled by advances in the preparation of specimens for APT [31], which has been increasingly used to study catalytically-active nanomaterials [32].

In search for quantification of Re-dopants in 2D nanosheets of molybdenum disulfide ($MoS_2$), they reported unexpected heavy elements like V and W in both Re-doped and non-doped nanosheets (Figure 2f-2h) [28]. In this case, the impurities likely originated from the commercial Mo precursor that contained 0.001 wt.% of miscellaneous 'heavy metals' impurities. In extractive metallurgy, separating Mo from V and W is difficult due to their similar chemical properties [33]. Additionally, also stemming from the Mo precursor, approximately 0.1 at.% of Na was measured, forming clusters of 4–5 nm in size, intercalated between the sheets.

In the following years, Kim and co-workers showcased multiple examples where APT revealed unexpected impurities in nanomaterials. Carbon-based molecules (*e.g.* ceterimonium) from synthesis on the surface of twinned Pd nanoparticles [29] were found to provide remarkable stability against oxidative dissolution. Additionally, in a different nanoparticle system, surfactant contamination was found on the surfaces of nanoparticles despite a rigorous washing process intended to remove residual molecules. The heavy presence of residual surfactant molecules raises questions regarding the surface coverage of these molecules and whether they block the active sites necessary for (electro-)chemical reactions in nanoparticle applications (see Figure 3).

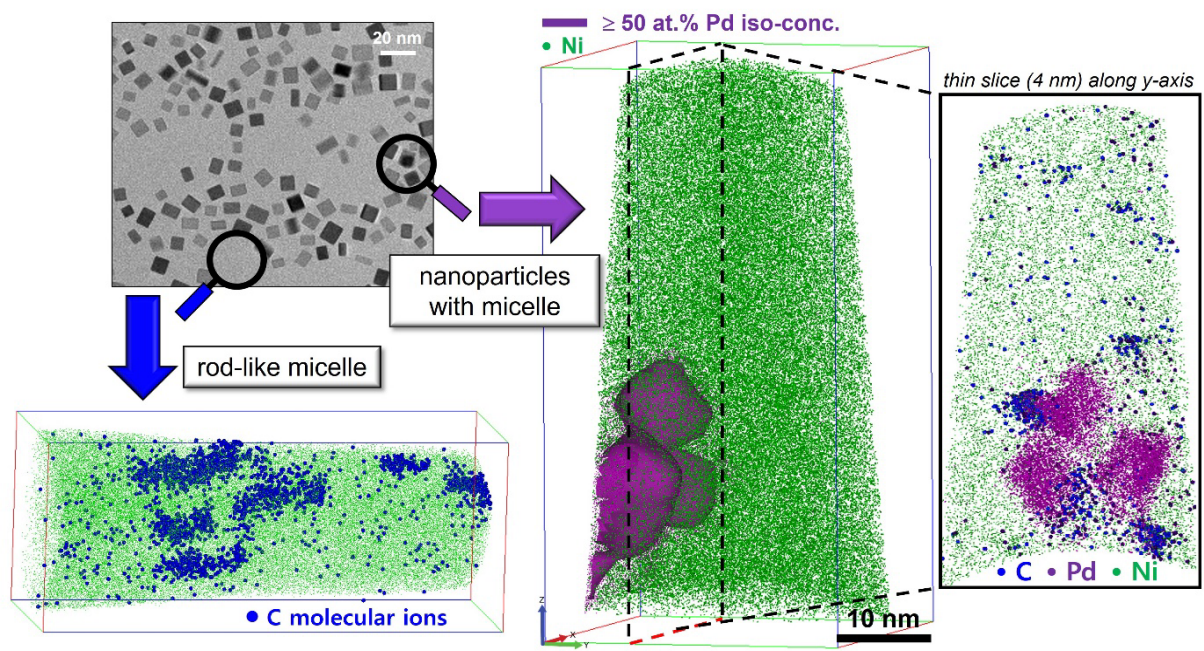

**Figure 3.** Remaining carbon molecules in a nanoparticle colloidal solution. Cuboidal Pd nanoparticles were synthesized following Ref [34]. The rod-like micelle and Pd nanoparticles were measured. Note that surfactant PVP molecules were attached at the corner of Pd nanoparticles [31].

## 3.2 From impurity to doping

Kim et al. then focused on wet-chemical synthesis using sodium borohydride (NaBH$_4$), a well-known reducing agent in both laboratory and industrial nanomaterial synthesis, for metallic nano-aerogels, which have emerged as a novel class of self-supported, porous materials with significant potential in electrocatalysis, including for the oxygen reduction reaction, glucose oxidation reaction, and alcohol oxidation reaction [35,36]. The introduction of NaBH$_4$ into the solution rapidly generates metal atoms and leads to a high concentration of crystal nuclei, which grow and merge into a metal nano-aerogel, as shown in Figure 4a-b. The 3D atom map resulting from APT analysis, in Figure 4c, evidences the complex morphology similar to electron micrographs of Pd nano-aerogels containing numerous interfaces, *i.e.*, grain boundaries [37] visible in Figure 4d. A composition profile measured along grain boundaries shows that the grain boundary contains 0.5 at.% Na and 2.3 at.% K. These unexpected impurities likely stem from the reducing agent (NaBH$_4$) and the Pd precursor (K$_2$PdCl$_4$), and their role in interacting with growing nanocrystals is often overlooked in nanoparticle growth models. Density-functional theory calculations revealed a strong binding energy of K and Na on Pd, suggesting that the surfaces of Pd nanostructures may be stabilized by alkali atoms binding to the surface during the initial stages of nucleation, but are detrimental for the cohesion of the grain boundaries.

Removing reducing agent would hinder the ease of production of nano-aerogels meaning that there is an unavoidable inverse relationship between the integration of impurities and the efficient production of nano-aerogels. Therefore, achieving an optimal balance in using or removing alkali atoms at grain boundaries may require a delicate balance for successful grain boundary engineering in nanoparticle synthesis. This insight challenges the conventional understanding of alkali metal presence in such nanoparticle systems, suggesting a deeper integration into the nanostructure.

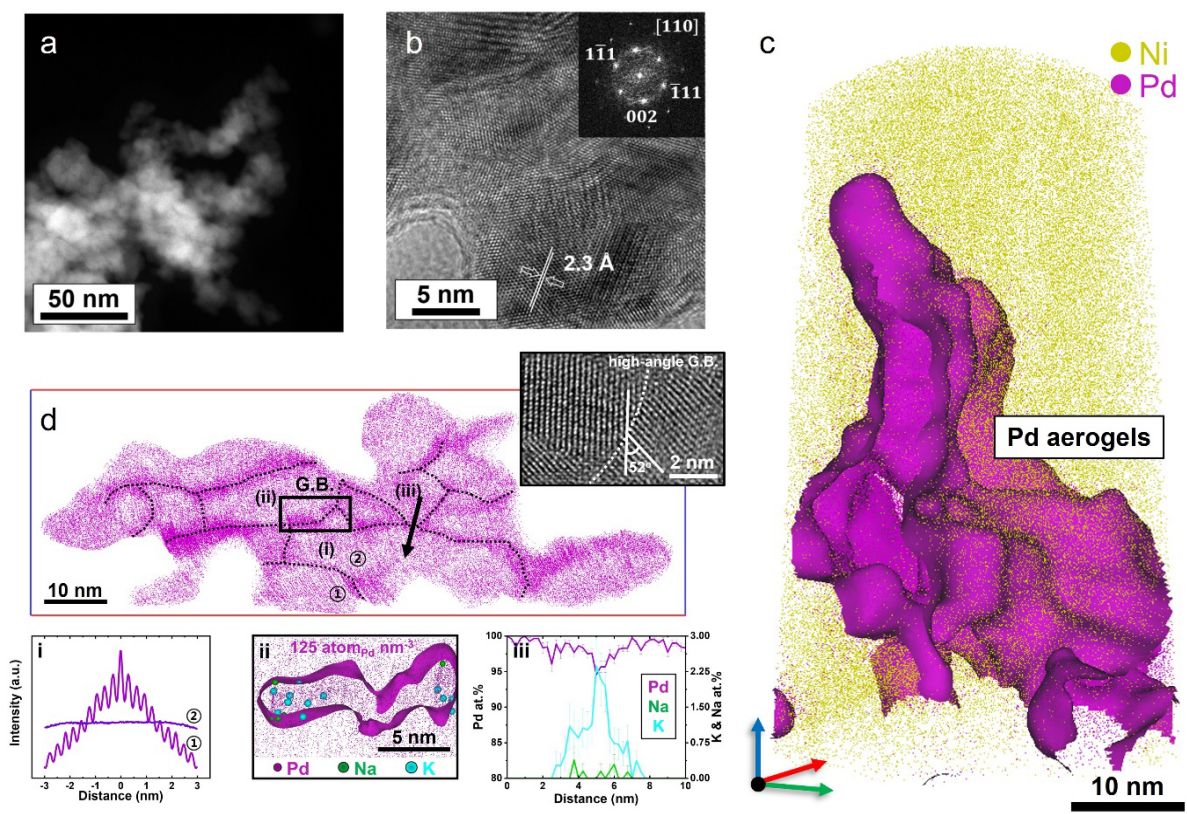

**Figure 4.** Advanced Characterization of Pd nanoaerogels. EM (a,b) and APT (c) images of as-synthesized Pd nanoaerogels. (d) 1-nm tomographic cross section of the aerogel with Na and K alkali atoms segregation along the grain boundaries. (Individual subfigures are from Ref. [37] and [38]).

Apart from alkali incorporation, Kim et al. reported ingress of B into Pd nano-aerogels [38]. In the first report of sodium borohydride reduction for synthetic chemistry, H. Brown et al. suggested that metal borides were formed as a by-product [39]. A lack of detailed characterization of B in nanoparticles resulting from $NaBH_4$ reduction, has led the assumption that particles are clean [35,36,40]. Figure 5a and 5b showcase that B is 10 times more concentrated than the alkali, and is homogeneously distributed within the nano-aerogels. Importantly, Kim et al. demonstrated that the B incorporation during synthesis could be controlled through adjusting the synthesis conditions including chemical potential (concentrations of the precursors), temperature, and injection speed of the reducing agent. The increased B content in solid solution was measurable by X-ray diffraction (Figure 5c), but the B remained distributed within the aerogel, as evident in Figure 5d.

In addition, they demonstrated that the ingress of B impurities could be turned into "doping", as the change in the electronic properties of the B-containing Pd leads to modifications of the hydrogen absorption and adsorption on the surface, which normally hinder the use of Pd as a catalyst for the hydrogen oxidation reaction (HOR). Through control of the synthesis conditions and resulting B-doping level, the catalytic efficiency could be optimised.

This precise control over B-doping levels not only helps in understanding the role of impurities in nanoparticle synthesis but also in enhancing the efficiency of hydrogen evolution reactions in alkaline conditions. Increasing B-doping in Pd nano-aerogels improves their catalytic performance, turning a typically undesirable impurity ingress into a controlled, beneficial doping process.

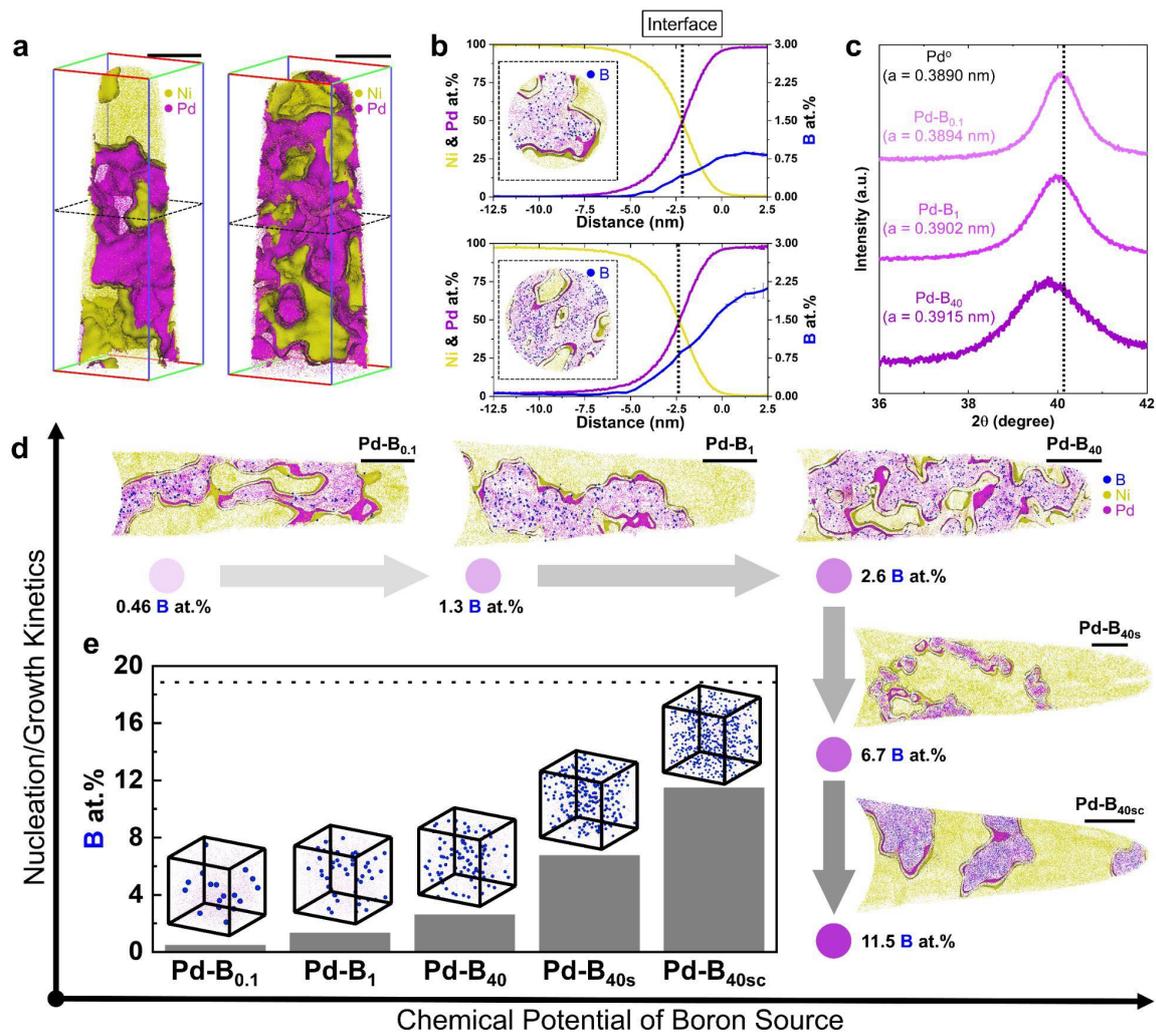

**Figure 5.** Impurity engineering in electrochemical catalysts. (a) 3D atom maps of B-doped Pd nano-aerogels with (left) low and (right) high amount of reducing agent during the synthesis. (b) 1D compositional profile along the surface for (top) low and (bottom) high amount of B. Each inset shows the corresponding top-view tomogram. (c) XRD patterns of Pd-$B_x$ samples synthesized with different reductant concentrations (x indicates the mole ratio between reducing agent to precursor). The dotted line represents the (111) peak position. (d) Schematic plot of chemical potential of boron source vs. Pd-$B_x$ nanoparticle formation kinetics. (e) B atomic compositions of Pd-$B_{0.1}$, Pd-$B_1$, Pd-$B_{40}$, Pd-$B_{40s}$, and Pd-$B_{40sc}$ samples (s and sc indicates the slow injection rate and cold temperature, respectively). All scale bars are 20 nm. Retrieved from Ref. [38].

To facilitate potential application of impurity-engineered nano-catalysts, the stability of dopant in nanocatalysts in operation conditions must be assessed, as a performance drop could prevent upscaling and commercialization for *e.g.* fuel cell technologies [41–43]. Yoo et al. used APT to rationalise the gradual decrease in activity HOR performance of Pd-B nanoaerogels during long-term cycling [44], Figure 6a and 6b. B is found to have leached out during the reaction, and the remaining B inside Pd had clustered. *Ab-initio* calculations show that the high stability of subsurface B in Pd is substantially reduced when H is adsorbed/absorbed on the surface, favouring its departure from the host nanostructure in reaction conditions. Hence, the $H_2$ fuel itself stimulates the microstructural degradation of the electrocatalyst, leading to a drop in activity.

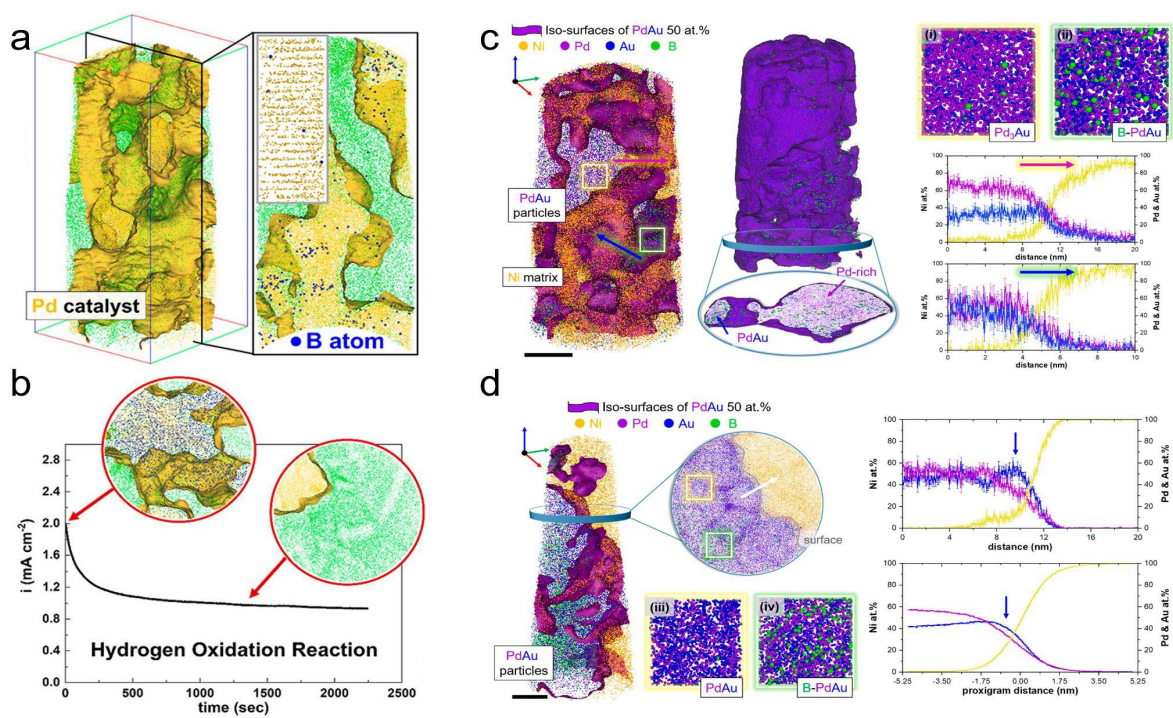

**Figure 6.** Pre/post electro-chemical test on the electrocatalysts. (a,b) APT results of pre/post-HOR Pd catalysts, respectively, and of (c,d) as-synthesized PdAu nanoparticles pre/post-ADT MOR, respectively. Scale bars for pre- and post-HOR 3D atom maps are 10 and 20 nm, respectively (Individual subfigures are from Ref. [44] and [45]).

## 3.3 Other functional nanomaterials

As discussed in the introduction, we drew from our recent experience, but many of these concepts are not new – they may be to certain readers or at least their importance might be. So now are these only applicable to monometallic nanoparticles or nanoaerogels? The methodology we have discussed above can be deployed onto more compositionally complex systems.

Aota et al. [45] investigated PdAu electro-catalysts, synthesized using the sodium borohydride method. These catalysts are efficienct for the methanol oxidation reaction (MOR) in alcohol fuel cells [46,47]. However, the microstructural origins of their performance degradation in service conditions remain unknown, limiting the ability to model their evolution and predict operational lifetimes. The pristine PdAu aerogel contains a mixture of $Pd_3Au$ and PdAu, as shown in Figure 6c, in which 0.2 and 2 at.% B are found, respectively. This tenfold difference in B concentration is attributed to difference in the interstitial sites in the lattice [45]. Additionally, after 1000 Accelerated Degradation Test-Methanol Oxidation Reaction (ADT-MOR) cycles, APT revealed Pd leaching from PdAu particles, leading to a transition from $Pd_3Au$ to PdAu solid solutions,. More importantly, it triggered the formation of Au-rich regions on the catalyst surface, and B leaching from both the surface and interior of the original $Pd_3Au$ regions. These findings suggest that B may play a crucial role in enhancing MOR activity. Therefore, understanding the mechanistic effect of B in Pd-Au catalysts and preventing surface Pd and B dissolution during MOR is vital for designing superior catalysts. These insights highlight the importance of understanding the chemical modifications occurring upon MOR to design new catalysts by aids of atom probing techniques.

Amongst other examples, we discussed nanowires and nanosheets above already, and we recently performed APT on MXenes, a family of 2D transition metal carbides and nitrides [48]. These materials are intensively investigated for various applications, including batteries and catalysis, and APT was able to detect Li and Na as intercalated species in addition to O, Cl, and F as surface terminations, as summarized in Figure 7. These elements originating from the wet-chemical synthesis all have an impact on properties, whether for later Li-intercalation or for catalytic activities, and can also play a role on their degradation, as the alkali elements have concentrated in TiO2 nanowires resulting from MXene oxidation, Figure 6c. The compositional space available for MXenes is enormous, and there are additional efforts to dope them, but targeted property optimization requires a deep understanding of how these incorporated elements from synthesis impact the solubility or distribution of other elements, in order to leverage them or define strategies for circumventing their detrimental effects.

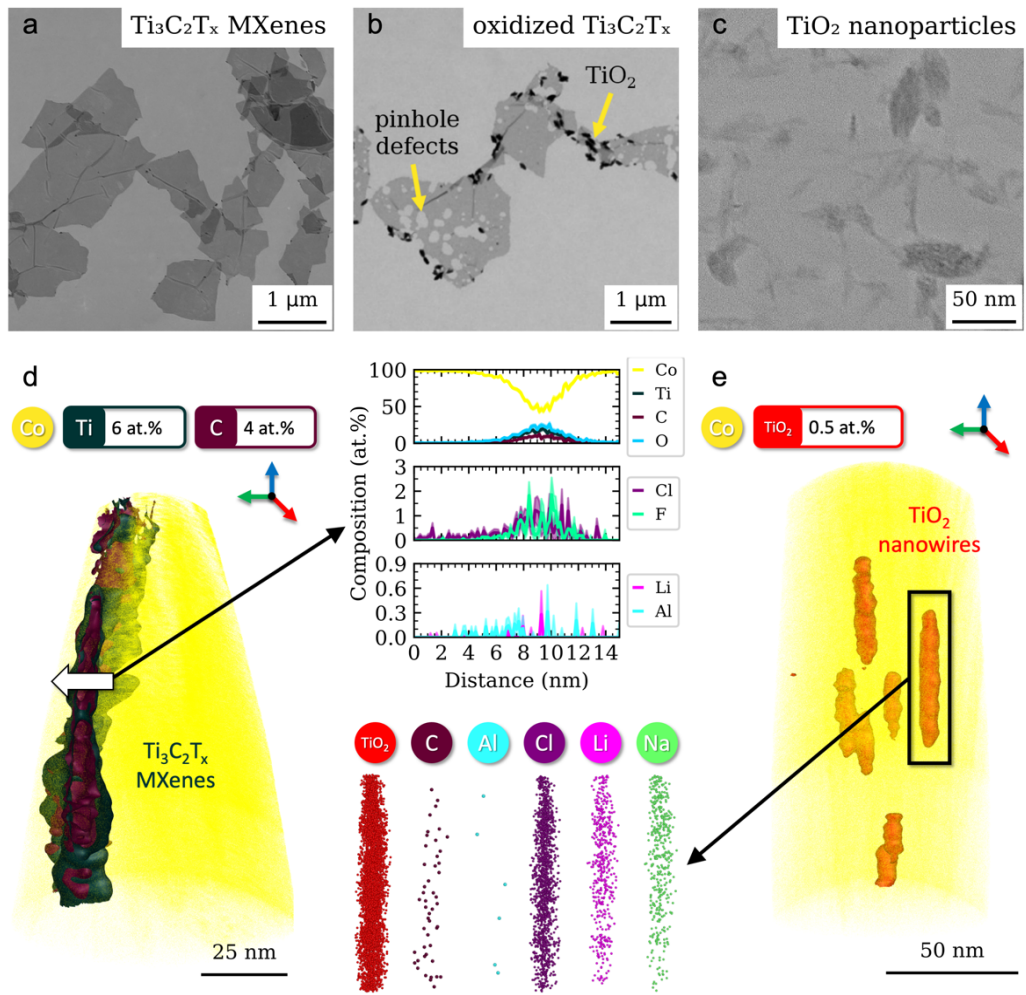

**Figure 7.** Caption: Characterization of the oxidation of MXenes. (a) As-synthesized Ti3C2Tx MXene nanosheets. (b) Oxidized Ti3C2Tx Mxene, stored in colloidal solution. (c) TiO2 nanoparticles resulting from Mxene oxidation. (d) APT analysis of as-synthesized Ti3C2Tx Mxenes, showing a 1D compositional profile across the agglomerated MXenes. (e) APT analysis of oxidized Ti3C2Tx Mxenes with elemental distribution maps, revealing the formation of TiO2 nanowires. The figure was redesigned based on the results in Ref. [48].

In bulk functional materials, similar concepts also apply. We recently focused on Heusler and half-Heusler compounds for their potential as thermoelectric materials [49,50], with much room for property optimisation. In the Fe$_2$AlV full Heusler, made of earth-abundant, non-toxic elements, for possible upscaling of thermoelectric technology, we demonstrated ingress of N during synthesis [51], leading to the precipitation of nitrides. We turned this observation to our advantage by introducing N2 into the atmosphere during processing in order to increase the density of nitrides [52], that helped scatter phonons and hence improved the thermoelectric performance [53]. We could also demonstrate that the expected doping through dispersion of Pt into the half-Heusler NbCoSn [54], in fact led to the formation of a phase along grain boundaries that offered a low-resistivity path, thereby leading to enhanced thermoelectric performance [55]. For thermoelectrics, the effect of impurities or dopants and their interactions with microstructural features is currently an intense area of research [56,57].

# 4 Perspectives

As a result of advanced characterization, we could evidence that materials systematically contain impurities. The presence of impurities can be either positive or negative, depending on the intended application – do they cause of accelerate corrosion or degradation? Can they be turned into an advantage and, through controlled synthesis conditions, dope the material and help to tune the material properties? This latter concept was sometimes termed "impurity engineering" [58]. This context brings us back to a perspective article titled "Will any crap we put into graphene increase its electrolytic effect?" [59] where the authors discovered that graphene composited with bird dejections made it more electrocatalytically active towards the hydrogen evolution reaction and oxygen reduction reaction. What is really contributing to the improved performance? Understanding the concrete chemistry-property relationship requires detailed studies, including high-end characterisation combined with fundamental atomistic simulations. In this case, the chemical analysis of bird drops reveals Fe and Mn impurities, along with trace amounts of heavy metals, and C, N, S, and P expected from organic compounds.

So far, we have not even considered incorporated H. H is ubiquitous, known to affect the properties of most materials, but is a challenge for quantitative microscopy and microanalysis. Recent progress in APT [60], makes quantitative assessment of how H affects structural [61] and functional[62] properties achievable in principle. However, there are still numerous challenges associated to the preparation of specimens [63], and their handling, as they often require cryogenic preparation and transfer [64], which has motivated novel frontier developments in terms of methodology and instrumentation [65,66].

To conclude, understanding the influence of residual impurities will become even more crucial to accelerate the optimisation of industrial processes, as the proportion of recycled materials in the feedstock of raw materials will keep increasing in the foreseeable future [67], in order to reduce the carbon emission associated to materials fabrication. This reinforces the need to consider careful material characterization to guide material design. Even though it may seem to make research slower, it helps deepen the understanding of property- or life-limiting microstructural features, helping to target the optimisation of the material's synthesis and processing for enhanced activity or operational lifetime, and hence sustainability. Smarter synthesis can avoid *e.g.* further alloying or doping, or heat treatment, which can save time and reduce costs.

If the difference in scale between these microstructural features and a full device can seem difficult to reconcile, dedicated workflows that involve arrays of microscopy and microanalysis techniques are already established and will keep being developed as techniques progress in the decade to come.

## Acknowledgements

We thank Uwe Tezins, Christian Broß and Andreas Sturm for their support to the FIB and APT facilities at MPIE. We are grateful for the financial support from the BMBF via the project UGSLIT and the Max-Planck Gesellschaft via the Laplace project. S.-H.K. and B.G. acknowledge financial support from the ERC-CoG-SHINE-771602.

On behalf of all authors, the corresponding author states that there is no conflict of interest.